# (Un)making AI Magic: a Design Taxonomy

Maria Luce Lupetti
m.l.lupetti@tudelft.nl
Delft University of Technology
Delft, The Netherlands

Dave Murray-Rust
d.s.murray-rust@tudelft.nl
Delft University of Technology
Delft, The Netherlands

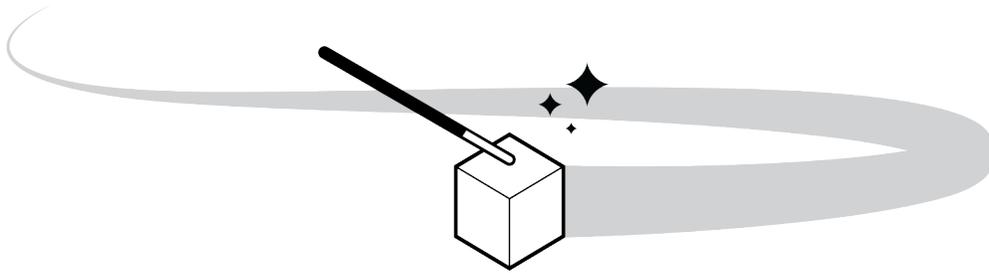

Figure 1: (un)making magic.

## ABSTRACT

This paper examines the role that enchantment plays in the design of AI things by constructing a taxonomy of design approaches that increase or decrease the perception of magic and enchantment. We start from the design discourse surrounding recent developments in AI technologies, highlighting specific interaction qualities such as algorithmic uncertainties and errors and articulating relations to the rhetoric of magic and supernatural thinking. Through analyzing and reflecting upon 52 students' design projects from two editions of a Master course in design and AI, we identify seven design principles and unpack the effects of each in terms of enchantment and disenchantment. We conclude by articulating ways in which this taxonomy can be approached and appropriated by design/HCI practitioners, especially to support exploration and reflexivity.

## 1 INTRODUCTION

*"The new AI can write and talk! It can draw, do fake photos and even make video! It even has AI folklore. Authentic little myths. Legendry."* [114]

As the science fiction writer Bruce Sterling emphasizes, there is a discourse of great excitement and wonder about all the novel artificial intelligence (AI) algorithms and applications that have been launched in the last years – a *Mardi Gras of AI*[114]. After the last winter, the current spring [43] of AI tools is now being leveraged to speed up and enhance a multitude of works, especially in creative practices where applications range from idea exploration [22, 66, 68] and ideation [128] to project documentation [21], creative partnership [71, 86], and so on.

AI has always been a metaphor-driven field [2], and the search for useful metaphors that articulate the possibilities of technology has a hand in shaping the field [7, 65]. As these tools are adopted, along with wonder and excitement, we see a metaphorical vocabulary that stems from the world of the supernatural and emphasizes the 'magical and enchanting' nature of these technologies. As an illustration, Derczynski [31] reports on a series of recurring terms used by a variety of professionals that show this metaphorical link to magic, such as *spellcasting* for providing or developing an input to a prompted language model; *alchemy*, to describe a practice of constructing model inputs to discover the different model outputs they lead to; *invoking*, to change the characteristics of a model output; and more.

Such a tendency to use the language of magic is not new – "it is often the case that new technologies are presented as magical" [18]. However, around AI tendencies towards magical thinking are emphasized as computer programs are seen to carry out tasks that were once unimaginable for computers to do [108]; to have skills that were once thought to be exclusive of humans [87], e.g., to converse [57] and express creativity [38, 76, 88]. The term artificial intelligence – particularly in contrast to alternatives such as 'complex information processing systems' [61] – invites us to attribute typically human properties such as "thought, imagination, memory, will, sensation, perception, belief, desire, intention, or feelings" to AI systems and products in which they are embedded [24]. Such emphasis on replication of human intelligence, and the ease with which we over-attribute abilities to computational systems, means that AI has now "come to have overtones of trickery" [125] as we discuss for example the gap between being able to produce sentences and actually understanding language [9].

Here, we work with the idea of *enchantment*: "the experience of being caught up and carried away", where our attention is captured, we are "charmed and disturbed", and our "background sense of order has flown out the door" [12]. Contemporary AI algorithms, especially deep learning, have a particular power to enchant. They embody a level of complexity that is hard for experts – let alone the general public – to grasp [18]. This complexity renders AI algorithms inherently opaque [5, 74, 96], in their structure, their internal patterns, and their links to the world [39]: "when a computer learns and consequently builds its own representation of a classification decision, it does so without regard for human comprehension" [16]. This gap of comprehension is the space through which magic flows [4], opening space for ambiguity and unexpectedness, and leading to truly enchanting experiences [85].

The language of magic and superhuman abilities is purposefully embraced by tech industries to build a public imaginary of AI as a silver bullet [73, 122] for solving all problems. As some examples: Future Tools [118] *"collects & organizes all the best AI tools so you too can become superhuman!"*; Google's new text feature for the Chrome browser is called *Magical* [81]; Figma's AI toolset is called *Magician* [42]: "Every little thing it does is magic. A magical design tool for Figma powered by AI."; and Runway AI [106] introduces its AI tools with the header: *"AI Magic Tools. Dozens of creative tools to ideate, generate and edit content like never before"*. However, as well as generating "interest in the field, spur[ing] financial investment, and trigger[ing] research and development", this language also contributes to obfuscating the actual practices involved in the 'making' of AI systems [36]; it disproportionately emphasizes the possibilities and opportunities offered by AI rather than its actual functionalities, blurring the line between the fantasy and the reality of these technologies [36]. Designing and communicating AI things as supernatural–enchanted–products, in fact, shapes the social perceptions of these systems, taking them out of the realm of mere technical tools to be regarded as socially capable agents [113] and/or socially valuable applications [18]. Building on Bennett's development of enchantment [12], Campolo and Crawford [18] define this phenomenon as *enchanted determinism*: "a discourse that presents AI, and deep learning techniques especially, as magical, thus outside the scope of present scientific knowledge, yet also deterministic, in the sense that AI algorithms can nonetheless detect patterns that give unprecedented access to people's identities, emotions, and social character". Working in this magical domain allows designers to focus on mastery of the illusions that they create [113] while minimizing concern for the consequences they cause [18, 36].

This paper investigates the ways in which design can contribute to increasing and decreasing the sense of enchantment in experiences with AI products, supporting an understanding of the dynamics and effects of magical thinking in the design of AI things. We do this by building a taxonomy of design principles that engage with magic and AI in the creation of interactive products. The taxonomy is based on a combination of theoretical development and critical reflection on two years of the *Interactive Technology Design* Master's course held at Delft University of Technology (TU Delft) between the years 2021 and 2023. While student work may not represent the full landscape and richness of AI things we can design, it gives a unique window into the conceptual development of designers who are engaging with AI. We reflected upon students' projects following the notions of enchantment and disenchantment, and used these to iterate on our taxonomy, developing a set of archetypal principles that contribute to creating AI-based products and services. We expand the discussion of these by articulating how designers can approach and appropriate the taxonomy in their practice.

## 2 MAGICAL THINKING IN THE DESIGN OF TECHNOLOGY

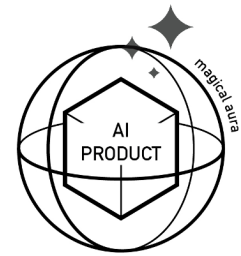

**Figure 2: AI products are inherently enchanting things as they hold a magical aura that is fed by AI uncertainty and errors. We define magical aura (the circles around the cube) in section 4, building on the work by Gell [51]**

> *"Any sufficiently advanced technology is indistinguishable from magic"* [26]

This famous quote summarizes a complex entanglement of phenomena at play when it comes to the perception of technology as a product of magic. Gell [51], committed to scaffolding the value and workings of art, provides an extensive explanation of how creative practices such as painting, poetry, and fiction can contribute to creating a sense of enchantment toward man-made things. But as Gell [51] further clarifies: "enchantment is immanent in all kinds of technical activity" and stems from the uncertainty we experience when encountering novel technologies about which we only have partial knowledge. Magical thinking and language have a long history in the design of technology and human-computer interactions (HCI). Designers and makers of technology are often regarded as the "magicians of the twenty-first century" [27]. Computational technologies are increasingly used to generate novel product configurations whose functionings cannot be fully understood [69, 104] generating a sense of wonder. These blend the familiarity of existing products with the unexpectedness and ambiguity of novel technologies [69, 104]. As Landin [69] explains, technological products can become magical as they bring unexpectedness and surprise to the interaction, while maintaining recognizable features.

In this section, we review two distinct approaches at the intersection of technology, HCI, design, and enchantment.

## 2.1 Designing for Enchantment

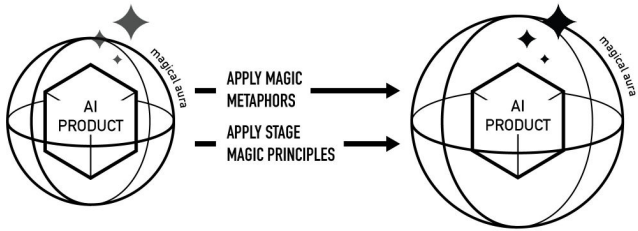

**Figure 3: Main actions identified in literature regarding the design of AI things that contribute to emphasizing AI's enchanting potential: 'Apply Magic Metaphors' and 'Apply Stage Magic Principles'**

> *"Magical technology consists of representing the technical domain in enchanted form"* [51]

Over the last decades, several authors have been exploring the parallels between HCI design and stage magic, such as "a need for consistency, the use of metaphors, and smoothness throughout the interactions" [4, 117]. This is particularly evident in those applications of computing where entertainment and engagement of the audience are key to a successful user experience, such as interactive performances, installations, and games [83]. Some metaphors from the world of magic have even gained so much popularity in HCI that they became emblems of interaction paradigms. The concept of a magic wand, and the related metaphor of spell-casting, is regularly used for embodying handled navigation systems that allow to control of smart environments [25], while the idea of magic mirror has become the recurring metaphorical reference for materializing the 'on-user' augmented reality paradigm [37, 80].

Magic, however, often materializes in HCI in a much more subtle way. When enhanced through the use of emerging technologies, familiar products become "more useful, more delightful, more informative, more sensate, more connected, more engaging" [104]. But together with becoming extraordinary, these products also become less easy for people to understand their inner functioning, creating a sense of wonder and enchantment. Different views do exist on what an object with the power to enchant is and does. Bennet [12] looks at enchantment more as a state in perception rather than as a quality of objects. Conversely, Rose [104] believes that an enchanted object is something that starts as an everyday object that gains 'magical powers' by the addition of technology, which makes the user both comfortable and captivating. Fisher (as in [84]), instead, also has a focus on enchantment as a quality of an object, yet this "does not remind us of anything we know and we find ourselves delaying in its presence for a time in which the mind does not move on by association to something else". Despite the different views, enchantment can be understood as a relational quality between user perception and the object's performativity. In particular, objects' ambiguity disorients but also engages, and enchants by offering the potential for unexpectedness and discovery [85]. This more implicit way of embedding magic in HCI, the design for enchantment, is actually very popular in a variety of HCI domains, such as the Internet of Things (IoT) and the human-robot interaction (HRI) fields where products are intentionally designed for encouraging people to "attribute properties such as thought, imagination, memory, will, sensation, perception, belief, desire, intention, or feelings to these products" [24]. Leveraging enchanting mechanisms (e.g., deceptive behaviors as in [79]) products are conceptualized and experienced as social agents rather than tools [24, 93, 113]. In this regard, Watson and colleagues [123] identified thirteen principles of stage magic that can be applied to user experience design, such as vanishing, transformation, and prediction.

Designing for enchantment, then, allows the achievement of seamless and engaging interactions with artificial agents [102], while also encouraging consideration of "aspects of interaction that often remain underdeveloped in HCI, such as cultural aspects of interaction" [105]. As Ylipulli and colleagues [129] argue "We deem that magic as a concept can mobilize thinking that helps to come up with idealized and alternative realities, and related ideas". The use of enchantment and 'make-believe' in technology design, however, is a double-edged sword. Many examples show how computational products easily turn from friendly enchanted things to spooky presences that surround us [32, 104]. Targeted advertisements on personal devices, for instance, are highly associated with feelings of creepiness [130], especially due to the partial inexplicability of how the systems work and the people's impression of devices being always listening [17].

## 2.2 Perils of Enchantment

The perceptual mechanism of enchantment itself–the gap in understanding that disorients and captivates– often plays a significant role in the key ethical challenges associated with digital products, especially when powered by AI. On the one hand, people have higher chances of improperly calibrating trust in artificial agents [30] if the capabilities of this are enhanced through the language of magic. As Wolf and colleagues [126] argue: "Users 'enchanted' by deceptive machines are likely to make inappropriate decisions based on these deceptions". Building on a classical quote from Plato [98], in fact, Turkle [116] argues that every deceiving thing can be also considered a form of enchantment and vice versa: everything that enchants also deceives. Deception, whether intentional or not, is inherently risky as it can give a false sense of mutuality in human-agent interaction (e.g., in sensitive care settings)), and it can be used to conceal non-anthropomorphic capacities (e.g., to hide a user monitoring mechanism present in a robot) making people less alert towards possible dangers, thus vulnerable [29]. On the other hand, the problem of enchanted technologies extends far beyond the issue of users' misunderstanding of product capabilities. Notions of magic and mechanisms of enchantments hinder the way for people to properly grasp the complexity of the ecosystems populated by technological products [82] and their risks [18].

> "The dominant narrative of technology as enchantment helps maintaining a cloak of impenetrability as to how technologies actually work, exploiting the genuine difficulty in grasping machine operations that run in the background, unseen, unheard, unknown, and incommensurable to human scale" [18]

With technological products becoming increasingly more complex and appearing as "magical unknowns" [4], enchantment also becomes a mechanism for masking exploitative work ethics [27, 36], power struggles [111], and dramatic environmental impacts [82, 112]. This calls for more cautious approaches to the design and communication of advanced technologies, especially AI for which it has become harder and harder to separate actual capabilities from industry and media promises [125]. This contributes to a phenomenon that Campolo and Crawford [18] define as *enchanted determinism*: "a discourse that presents AI, and deep learning techniques especially, as magical, thus outside the scope of present scientific knowledge, yet also deterministic, in the sense that AI algorithms can nonetheless detect patterns that give unprecedented access to people's identities, emotions, and social character".

## 2.3 Designing for Disenchantment

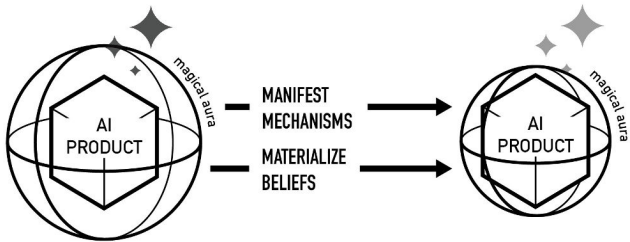

**Figure 4: Design actions that contribute to reducing AI products' magical aura and diminishing their enchanting potential**

The more AI products take part in our everyday practices, the more it becomes of crucial importance to "distinguish research, pouring from companies and laboratories, from speculation, fantasy, and fiction" [125] and to address social and ecological implications along with technical discourses [101]. Research in the area of explainable AI and AI auditing implicitly sets out to answer these needs. As Campolo and Crawford [18] explain, experts in the area of AI interpretability in machine learning are developing ways to provide rational explanations to classifications and predictions that enchanted discourses of AI withhold. AI auditing, instead, looks at AI algorithms documentation and related operations to account for and ensure the AI system's trustworthiness [67] as opposed to the dominant claims of progress and liberation. These approaches represent important strategies for experts in various domains, from technical to legal, to *disenchant* AI, as they allow for "dissecting and inspecting" algorithms and the systems surrounding them. However, their effect on the way AI products are shaped, communicated, and perceived by the public is indirect: it is up to designers and developers to take on the responsibility for the rhetorics that AI is presented with and the semantics of how it is experienced.

Conversely, a distinctively designerly perspective comes from the field of Critical and Speculative design in which *provocative designs are used to bring awareness to the myths and beliefs we bestow on AI* things and technology more broadly [33, 77] as well as to *manifest often obscure mechanisms* of technology. For instance, allegorical steering wheels have been used to challenge the myth of the autonomous car and its related narratives [77], and custom PCBs satirically intended to predict luck and harmony, have been designed to actually make us reflect on the beliefs and assumptions of superhuman capabilities and objectivity that we project on AI [41]. With complex technological infrastructures, a range of design practices needs to be brought together to articulate the multiple mechanisms at play [89]. Critical and Speculative projects, however, often remain marginal practices, and their effects are limited to public debate, rather than getting embedded in actual product development processes [127].

## 3 TOWARDS A TAXONOMY OF MAGIC IN AI DESIGN

As critical design and HCI researchers, we share the view of the many authors [18, 36, 113, 125] that problematize the use of the magic rhetoric concerning novel technologies, and especially AI systems. We agree with Sharkey and Sharkey's [113] position that it is a moral imperative for designers of AI systems to be truthful about the actual capabilities of the technologies they are developing, along with being honest about the intent of creating the *illusion* of intelligence. Fanciful construction of AI as magic, as holding higher-than-human capabilities, while disregarding its actual feasibility would be a great mistake [92]: being swept up in enchantment can blind one from the actual implications of AI, and prevent the development of strategic methodological sensitivities that critically ground AI analysis and claims [36]. However, as previously discussed (Section 2.1), we also see the pull of enchantment around new technologies, and argue that rather than dismissing it altogether, we should build a deeper understanding of its dynamics and how to skillfully and honestly leverage the possibilities of enchantment in the design of future technological products.

More ontological resources are needed to navigate this intricate space [23, 75], and theoretical tools such as taxonomies hold great potential. In this case, we set out to create a taxonomy that can not only help structure the body of knowledge in the area of magic and AI and predict its future developments [54] but also provide design/HCI practitioners, educators, and researchers with a resource for informing future designs and for practicing reflexivity on their processes.

In what follows, we illustrate the process and results of our *(Un)making AI Magic* taxonomy development. We focus specifically on AI technologies not only because these are considered the most important of the fourth industrial revolution [53] but also and foremost because of the strong connection between AI and the narrative of technology as magic [72, 92, 107]. We aim to tease out the mechanisms for enchantment – and disenchantment – that designers can draw on around AI systems.

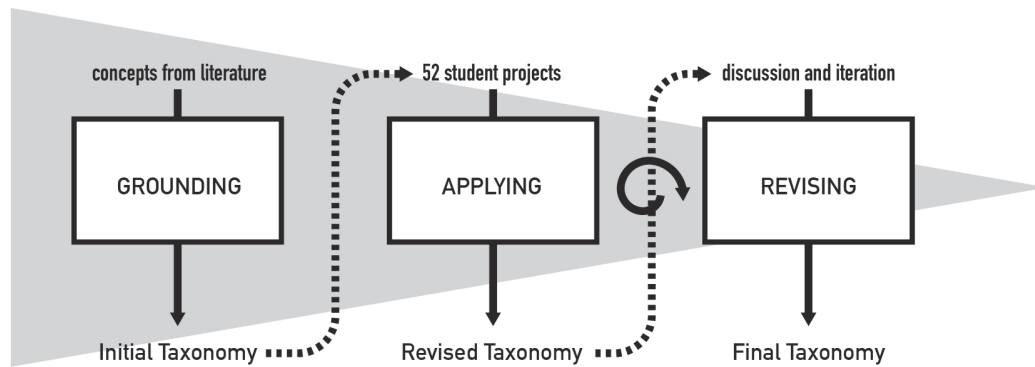

Figure 5: Taxonomy building process: grounding, applying, and articulating.

This does not mean eliminating the sense of enchantment entirely: as Sharkey and Sharkey [113] discuss, just as cartoon animators do not need to sell their characters as truly thinking beings, AI designers can leverage the active participation of people and their willing suspension of disbelief to maintain a sense of enchantment without deception. Our taxonomy builds on Gell's "halo-effect" of technology – the *inherent property of the technological product to enchant*. This halo is intrinsically bound to the beholder's mental model of the technology in use, as it responds to the uncertainty we experience when encountering novel technologies about which we only have partial knowledge [51]. We focus on strategies that designers can employ to affect the halo–the perceived *magical aura* of technological artifacts.

## 3.1 Taxonomy Building Approach

Following [94, 95], we employed a "conceptual–to–empirical" approach to building the taxonomy. This involves setting out the characteristics of the taxonomy, defining a stopping condition, and then a process of iteration where an initial conceptual version of the taxonomy is grounded in a fixed set of examples. We first set our area of concern, or meta-characteristic, as the different strategies that can be employed to modulate the magical aura of interactive AI products: that is each dimension in our taxonomy will be something that can be done to affect the kind of enchantment that a product engages in. Our example set is a collection of 52 student projects (see Section 3.2). The stopping condition is that i) objectively, all projects are associated with at least one strategy, and ii) subjectively, all strategies of interest seen in the projects are accounted for.

After setting these conditions, the process involved three main steps, expanded in Sections 4–6:

- *Grounding* (Section 4). An initial deductive approach allowed us to derive the first set of categories from a combination of literature and personal reflexivity. This process started from intuition, theory, or conceptualization and identifies dimensions and characteristics in the taxonomy by a logical process derived from sound conceptual or theoretical foundations [95].

- *Applying and revising* (Section 5). Next, we empirically articulated the taxonomy through observations and reflections on the collection of 52 design student projects. Building on design academic traditions, here the empirical work was approached as an *annotative form of knowledge production* [50]. As with annotated portfolios [49], which are 'descriptive (of past occurrences) and intended to be generative-inspirational (of future possibility)' [15], we used projects' curation and critique to find patterns and to explicate how particular artifacts and their features embody elements and dynamics of a theoretical space [50]. This was carried out by the first author's examination of the projects through the classification identified in step 1 with the addition of missing categories to account for new discoveries.

- *Iteration, agreement and articulation* (Section 6). Through iteration and discussion with the second author, we verified the first author's classification and finalized the taxonomy.

## 3.2 Taxonomy Corpus

To develop the taxonomy, we reflected upon a collection of 52 design student projects, illustrated in Figures 7 and 8 that were the outcomes of a 20-week interaction prototyping course from two years (2022/23) of the *Interactive Technology Design* master course held at TU Delft. Students were explicitly focused on understanding and using novel AI tools in interaction design projects, developed through an iterative prototyping process. They were provided with tutorials and dedicated technical support for working with AI tools such as Teachable Machine, Edge Impulse, Voice Flow, and more, as well as with brief and provocative lectures on conceptual aspects of interaction design and AI. This represents a large body of work by developing designers, giving a window into their conceptual development. It illustrates how designers may be inclined to think about AI in magical terms and whether and how this may translate into specific product features and design dynamics.

Design education is often a testbed for the exploration of theories and methodologies in design [52, 90]. For instance, the notion of enchanted objects by Rose [104] was explored and articulated through his educational work at the Massachusetts Institute of Technology (MIT) where he was teaching the Tangible Media graduate course.

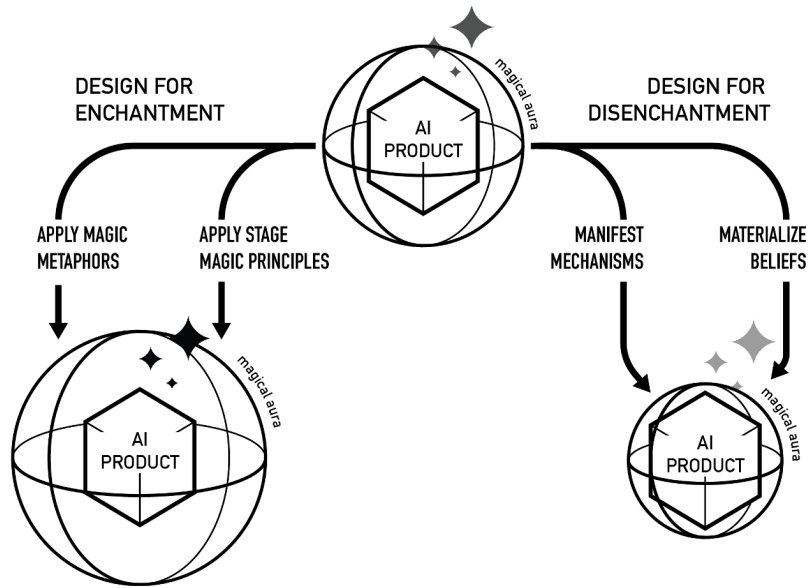

**Figure 6: Initial version of the (Un)making AI Magic taxonomy, based on literature, showing two approaches to designing for enchantment, through applying magic metaphors or stage magic principles, and two strategies for disenchantment: manifesting mechanisms and materializing beliefs.**

However, we do acknowledge that students' work may not represent the full landscape and richness of design in relation to magic and AI, and address this further in the discussion.

*3.2.1 Ethical approval and students' consent.* Students were asked to provide consent for using their educational outputs in research under university ethics board cases 2850 (2023) and 2251 (2022). In all cases, students were free to opt-out, and the consent was gathered by someone who was not part of the core teaching/marking team to avoid power imbalances. They were asked whether their concept cards (project images with descriptions) - which are the basis of this work - could be interpreted. We intentionally do not mention student names in the paper to avoid a sense of public critique of their work.

## 4 GROUNDING THE TAXONOMY

Starting from the focus on the *magical aura of AI technology* and building on the literature discussed in Section 2, we identified two broad categories of action between design and the magical aura of an AI product: enhancement and diminishment.

**Design for enchantment** covers actions that enhance this aura, making products feel more magical, building on Section 2.1, Figure 3. Concretely, *Borrowing magic metaphors* is an explicit approach, engaging with the language and imaginary of magic, such as when naming products (see many examples in Section 2.1). *Applying stage magic principles* works implicitly, to create seamless, fascinating, and engaging product experiences. For example, the NEST smart thermostat uses anticipation [62] to create a sense of magic functioning. **Design for disenchantment** conversely focuses on reducing the magical aura (Section 2.3, Figure 4). *Materializing beliefs* involves using provocative designs to bring awareness on the myths and assumptions we bestow to AI [59] or highlighting myths of AI objectivity about non-quantifiable things like luck [35]. *Manifesting mechanisms* and properties of AI products, such as the heavy hunger for data that AI systems come with [1], reduce the aura by making the mechanics more apparent to the end user.

This led to the initial structure seen in Figure 6.

## 5 APPLYING AND REVISING THE TAXONOMY

The students' projects (Figures 7 and 8), summarised through pictures and descriptions, were clustered and annotated by the first author following the four main design principles identified in the first version of the taxonomy (Figure 6): *apply magic metaphors* (Section 6.2), *apply stage magic principles* (6.1), *manifest mechanisms* (6.5), and *materialize beliefs* (6.4). To properly understand the projects, and especially the way AI was used and conceptualized, the analysis made use of various project materials that captured the students' process and outcomes, such as pictures, illustrations, videos, descriptions, and written reflections.

As a result of this activity, we found that a great number of projects embodied principles of stage magic (18), while some leveraged magic metaphors (4) and manifested mechanisms (3), and only one seemed to represent the principle of materializing beliefs. As a great number of projects (N 28) did not fit within our four initial categories, we iteratively revised the taxonomy (Figure 9) and identified three additional categories: *presume AI* (4), dispel enchantment through *play with AI* (1), and *summon AI as supernatural entity* (15). Finally, we excluded 5 projects from our classification and analysis,

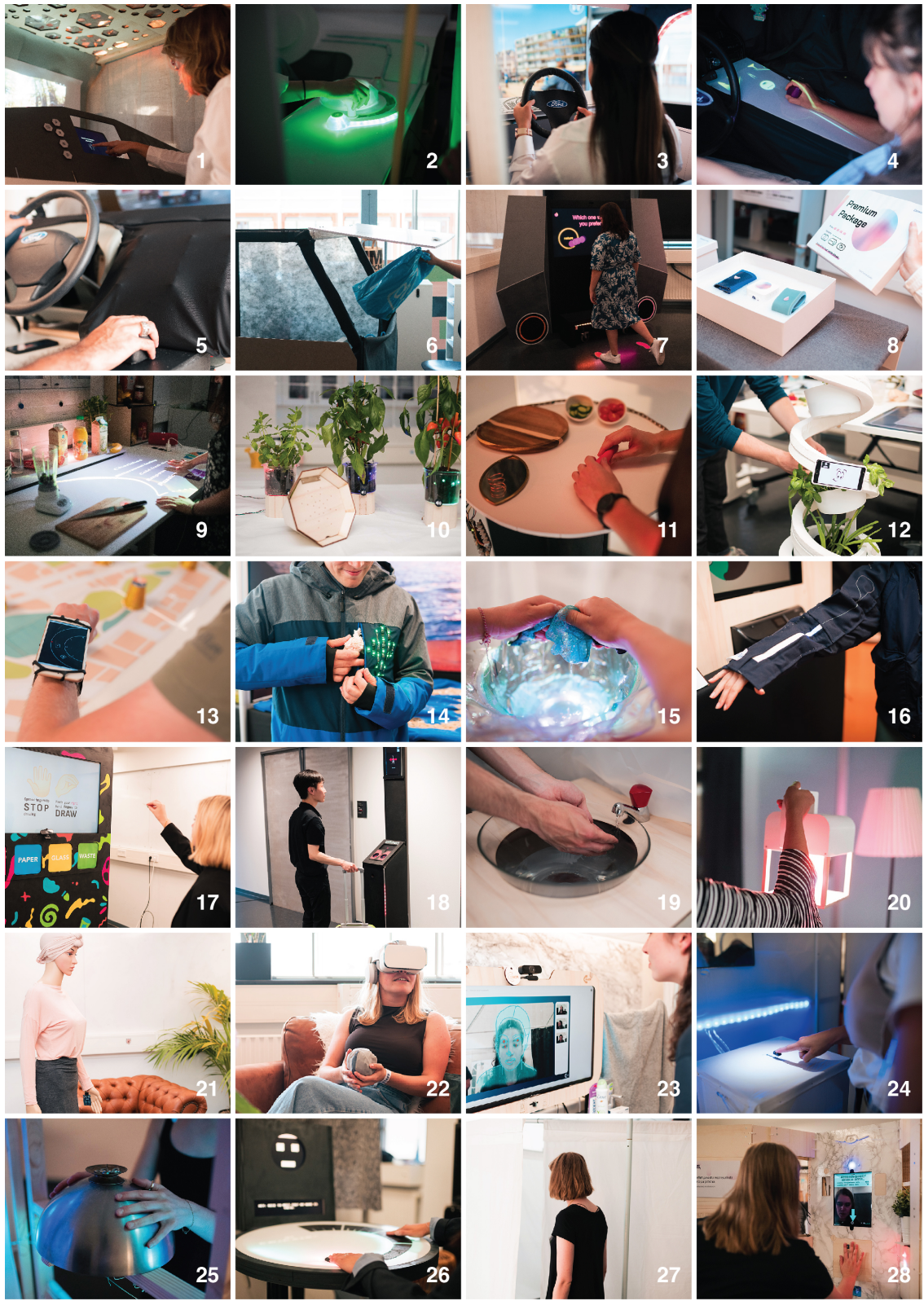

Figure 7: Overview of the 28 projects resulting from the 2022 edition of the Interactive Technology Design Course

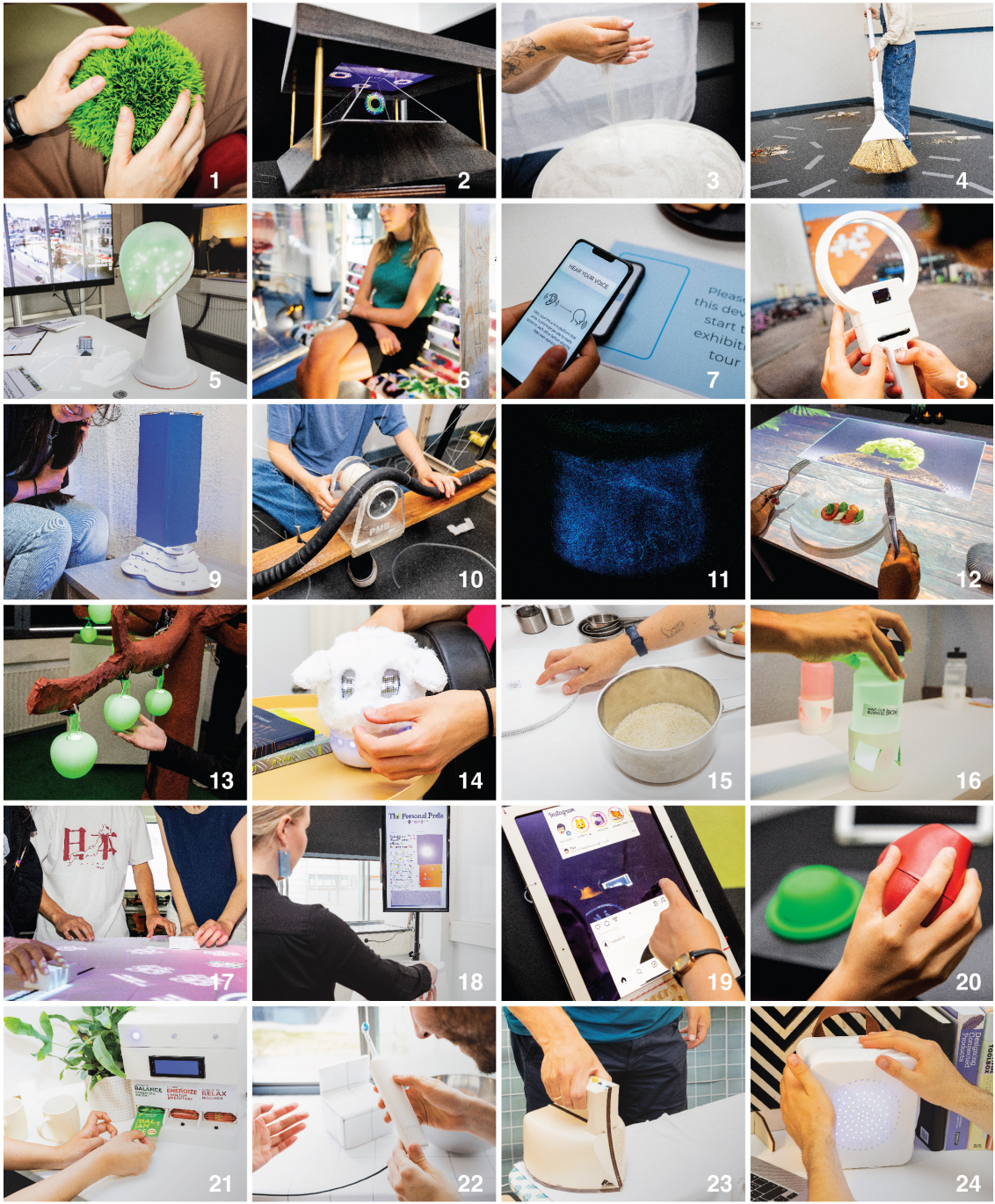

**Figure 8: Overview of the 24 projects resulting from the 2023 edition of the Interactive Technology Design Course**

as these did not engage with AI, either in the conceptualization or prototyping.

We note that the vast majority of projects operated within the space of enchantment, rather than disenchantment. This could be interpreted as a partial lack of critical perspectives in the design of AI things, or a result of the framing of the course. Rather than quantifying, however, our interest here is to learn what possibilities and dynamics each of the identified design principles affords and how. In the following, we unpack these discussing a selection of representative examples.

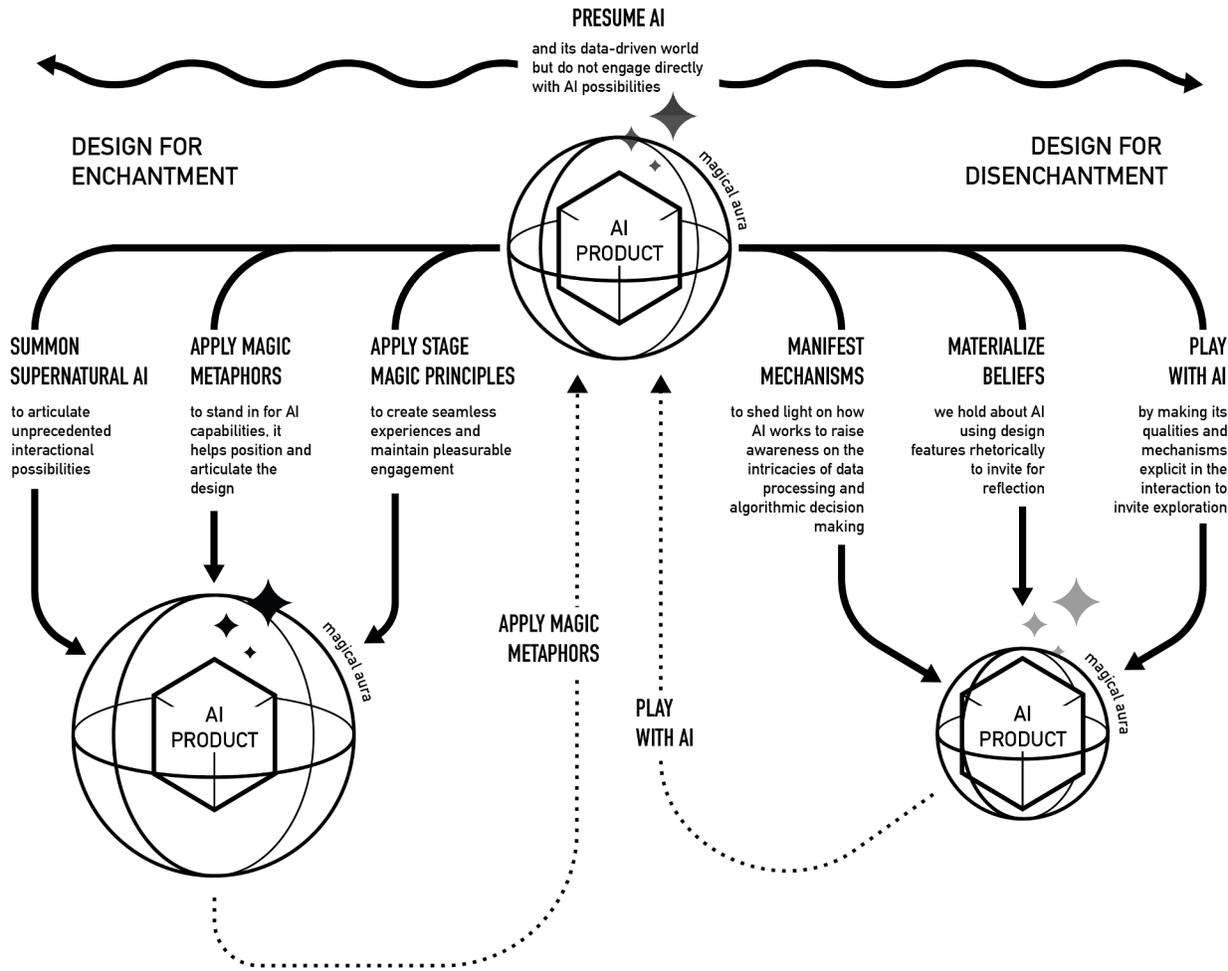

Figure 9: The revised version of the (Un)making AI Magic taxonomy from Figure 6 after the analysis of the projects uncovered the additional enchanting principle of summoning supernatural AI, possibilities around playing with AI, and a new category of projects that take AI as a given without directly engaging with it.

## 6 ARTICULATING THE TAXONOMY

Here we present our final taxonomy (Figure 9), through images and descriptions of representative projects.

### 6.1 Apply stage magic principles

Several projects (18) manifested an approach to design characterized by the use of stage magic principles to create engaging and seamless interactions. *Miron*, an iron that physically resists and opposes user intentions when ironing clothes, lets people experience the agency of products. *Summit of the Objects* takes this further to allow users to negotiate with objects, so that opening a tap for water begins an animated debate about how resources should be used, and which product should operate at a certain time. Projects in this category also create enhanced experiences of private and public environments: the *Fordy* car manifests its ability to anticipate needs by opening the rear door when a user approaches with shopping bags; *StairsOverElevatorUse* uses computer vision to understand when to nudge users to use the stairs instead of the elevator or not (e.g. when carrying luggage). In contrast to other categories, projects that apply stage magic principles present a relatively advanced implementation of technology, in a way that is so seamless that the technological mechanisms are not perceivable by the user. They create smooth and intuitive experiences that 'feel like magic'.

The *Under the Loop* project (Figure 10), for instance, consists of a device that resembles a magnifying glass. This allows citizens to provide input on future city developments, thus enabling administrations to learn about the thoughts and concerns of the public. The device is fitted with sensors (e.g., for measuring air quality) as well as a microphone, used to collect spoken comments about problematic situations that a citizen may encounter in their surrounding. A speech-to-text algorithm is used to translate the comments into data and, together with the sensor data, place these on a heat map of the city. Of all this technical functioning, the user only perceives the presence of the microphone and interacts with a button. The interaction with the device is an act of transformation: from a personal experience of space and place, spoken opinion is brought into a publicly accessible map of issues and contestation. As such, the designers engaged with key principles of stage magic: the changing state of the matter at hand [123] and the misdirection of attention [117]. Projects in this category, then, focus on promoting engagement and seamlessness [124] in use, and AI here mostly represents a tool for achieving unprecedented user experiences.

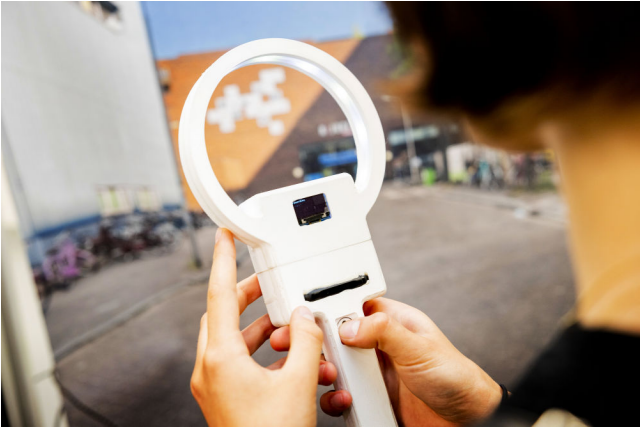

Figure 10: *Under the Loop*: citizen participation in placemaking using a magic magnifying glass to collect data and opinions

## 6.2 Applying Magic Metaphors

This category includes a few (4) projects, yet they are powerful. This category makes use of magic metaphors to describe and shape projects, and enhance the perception of the designed artifacts as magical. Our initial conception of this category was that the magic metaphors work as communication devices, aimed at supporting the end users' mental image of the AI product. However, through the reflexive activity of engaging with the projects, it became clear that the borrowing of magic metaphors is also a way for the designers themselves to unpack the features and interaction dynamics that an AI product may embed. The *LUMI* project (Figure 11) is designed to sensitize hotel guests about the scarcity of energy and promote sustainable behaviors. It very explicitly builds on Harry Potter's concept of *deluminator*: a magical lighter-like device that is used to absorb as well as return light from any light source to provide cover to the user. *LUMI* works under a similar principle: it is a lantern-like device, a single light point that holds light or sends it to different objects in a hotel room. In this case, while the metaphor may be known or not to the audience, it was strongly present in the imagination of all the students in the team who, through this, managed to agree on an overarching working principle, to envision the interaction architecture and components, and to build a coherent and understandable interactional aesthetics.

In a way, then, magic metaphors can help consciously guide the design of AI products as enchanted things, to bring focus on how magic principles may be embedded. These projects often leverage stage magic principles (Section 6.1): *LUMI* uses an optical illusion to give the impression that the same light 'jumps' from object to object.

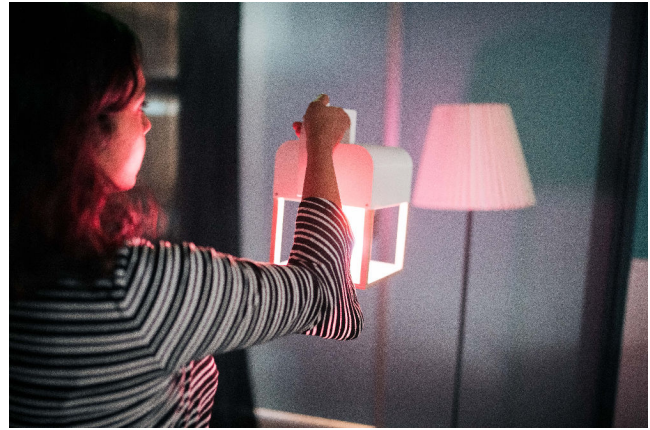

Figure 11: *LUMI*: embodying light energy through the magic metaphor of an enchanted lantern.

However, by explicitly using magic metaphors these projects emphasize and build on the supernatural conceptualization of the designed artifacts. This is a significant difference for the phenomenon of enchantment: as Tognazzini [117] explains, with stage magic spectators engage with an illusion and suspend their disbelief, going along with the idea that the *magician* is magical, even if only within the boundaries of the performance. AI products leveraging magic metaphors invite users to suspend their disbelief, making explicit the staging nature of the magic at play but centering the magic on the *object* itself, which both enchants and disenchants.

## 6.3 Summon AI as supernatural entity

The most prominent category that we identified in the attempt to understand the projects that did not fit with our initial classification is one that manifests a view of *AI as a supernatural entity*. This view is coherent with our general understanding of how the magical aura of AI influences designers' understanding of the technology they are starting to engage with. Through the analysis of projects, we further realized that this vision can also translate into an approach to the design of AI things themselves. A great number of projects (15), in fact, engages with AI in a fashion that sits in between the use of magic metaphors and the application of stage magic principles, yet has very distinct implications. AI is 'summoned' as an entity with supernatural powers, such as the ability to interpret and interfere with complex human experiences. In the *Reframe your Thought* project (Figure 12), AI is described as a tool that adds a visual dimension to therapy, understands conversations between therapists and clients, and generates positive images as alternatives to the ones representing the struggles of the person going through therapy. The underlying assumption is that AI is capable of understanding what human struggles are and what better alternatives there could be.

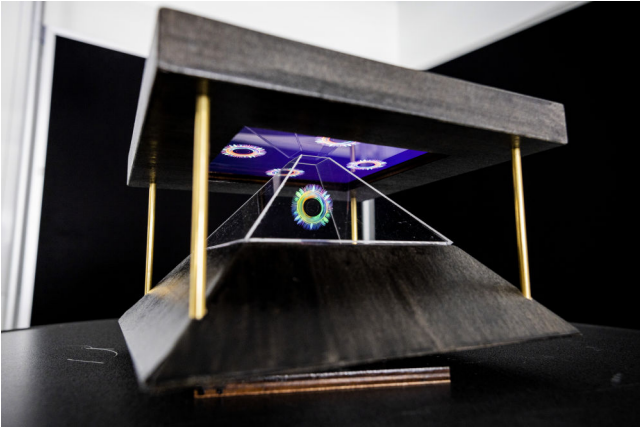

**Figure 12:** *Reframe your Thought*: generative AI techniques that create the illusion of a superintelligent being assisting in therapy.

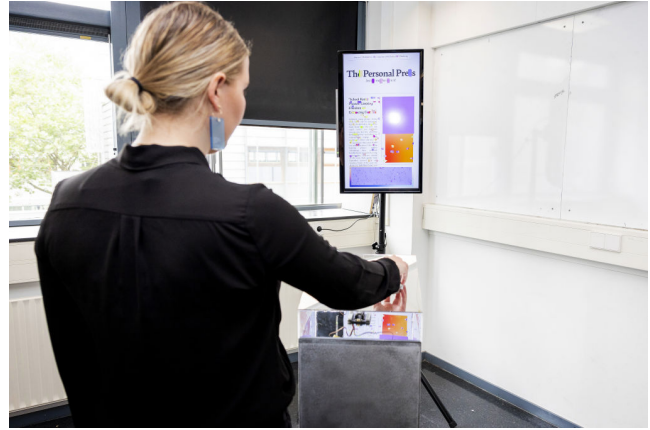

**Figure 13:** *Colored Realities*: a dynamic digital newspaper generated by large language models that allow users to control the political leanings and writing style of the text of any given news item.

The narrative, in these projects, is the one of AI as a possible superintelligence [14]. This is a controversial subject, and it is not uncommon for projects in this group to reveal a lack of thorough and critical engagement with the nature and qualities of AI. Despite the technical accomplishments of *Reframe your Thought*, in many projects seen here, storytelling and Wizard of Oz techniques are extensively leveraged to materialize the design concepts, often at the expense of technical development. The aesthetics of the projects also contribute to conveying this vision: AI things are shaped as abstract and symbolic objects, in which it is hard to determine where technology sits and how it works. An AI system may remind an Egyptian temple (*Reframe your Thought*), a shell (*The Shell*), a totem (*Connective Senses*), and even a sand garden (*Sand me Awareness*). Designers are caught themselves falling for the enchanting power of AI and 'dragging' the audience with them through the staging of the 'supernatural AI'.

### 6.4 Materialize Beliefs

This category covers projects where provocative artifacts are used to bring awareness to the myths and assumptions we bestow to AI. Somewhat surprisingly it includes only 1 project from the course, despite being a strong staple of critical practice (e.g. [28, 64]). While examples in the category *summoning AI as supernatural entity* (Section 6.3) can also be considered as implicitly materializing the belief that AI holds superior abilities, projects in this category take a critical stand to the beliefs articulated, exposing them for scrutiny. As we learned from the literature, critical and speculative designs can be used to address matters of concern regarding AI beliefs and intentionally promote discussion about these. In the *Colored Realities* project (Figure 13) this approach was translated into a generative newspaper, presented on a screen and controlled through a dial placed on a pedestal. The dial allows people to select a writing style or a political standpoint they would like to read their article with.

When turning the dial, the text in the article instantly changes to reflect the desired style. The project makes evident how using AI capabilities for generating personalized content can easily turn into an instrument of power. The generated articles are based on the same facts but leave out details in order to follow a specific style and set of values, obscuring other viewpoints and hindering people from getting the complete picture of events.

In contrast, while the project *Reframe your Thought* is grounded on the assumption that AI is capable of understanding human struggles and acting to promote positive change, this capability is not emphasized but rather quietly embedded into a 'gentle' product that mediates between a therapist and a client's conversations. A project designed for materializing beliefs [28, 59, 64], instead intentionally puts the spotlight on typically hidden features of the technology. These beliefs are translated into the product features - such as the dial in the Colored Realities project - inviting the audience to try out, indulge in experimentation, and see and feel what happens if we engage with a given belief. By prominently materializing beliefs, exaggerating and confronting them, this type of design disenchants.

### 6.5 Manifest mechanisms

This category covers projects that manifest the mechanisms of AI, as a way to disenchant. This is close to the approach of materializing beliefs (Section 6.4), but distinct in that here it is the mechanisms of action rather than beliefs and possibilities that surround them that are highlighted. The projects in this category (3) built seamless products and interactions that engage with popular concepts associated with AI, such as personalization (*Ready for your tea?*), social connectivity (*Own Faces*), and behavior change (*A Closer Look*). Together with providing an apparently clear and smooth interaction, the projects in this category also all present a 'backside' intended to put a spotlight on the hidden mechanisms at play and their potential intricacies. *A Closer Look* (Figure 14) is a smart health system in the form of a mirror that allows users to do daily checkups at home, as well as to get medications. The panel is equipped with several knobs and sliders that allow the user to select preferences and set parameters. However, together with the interface in the front, the mirror has knobs and sliders on the back too. As the piece is intended to be presented as an exhibition artifact, two people can experience the effect of shifting control between choices from the primary user in the front and the ones from the person hidden in the back. The last embodies the hidden mechanisms at play when we interact with AI-powered products that heavily leverage user profiling and provide info and recommendations based on hidden parameters.

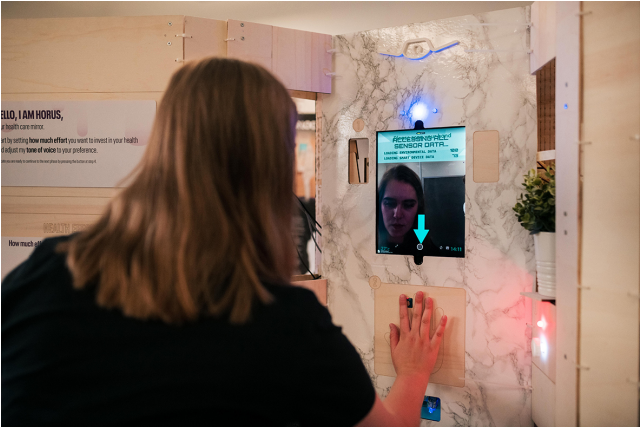

**Figure 14:** *A Closer Look*: a critical exploration of AI data collection and personal wellbeing, manifesting surveillance capitalism through a magic mirror, that invites visitors to play the role of either a client or the invisible data architectures that supply them with pharmaceuticals.

This design approach takes a distinctively critical stand and explicitly operates to disenchant the audience, in line with the idea from the literature that when we look inside the enchanted object and understand the inner workings we see that no magic exists [104]. Our reflexive analysis of the student projects allowed us to add a specific design principle that seems to be intuitive and effective for engaging with inner workings: the backside interface. The backside panel of *A Closer Look* project has some similarities with the backside screen display of the *Ready for your tea?* project in which user data is shown together with the overview of ongoing nudging activities, based on industry interests. In this project, in fact, the focus is on the intricacies of having smart products using nudging behaviors as they represent proxies for industries having at heart clear commercial interests, rather than the well-being of clients. In the *Own Faces* project, instead, the concept of a backside interface translated into a much more experiential design feature. The project, which revolves around the importance of image ownership when interacting with social media, invites the user to take a selfie in a photo booth and then post it online. In the moment the photo is taken frontally, other flashes appear from the back and sides, and multiple pictures end up being posted on social media. Through this, the experience translates the hidden working of image appropriation and manipulation that can happen online and makes it experiential and confronting. The concept of *backside interface*, then, is a useful way of manifesting mechanisms that can translate into diverse forms of design.

### 6.6 Play with AI

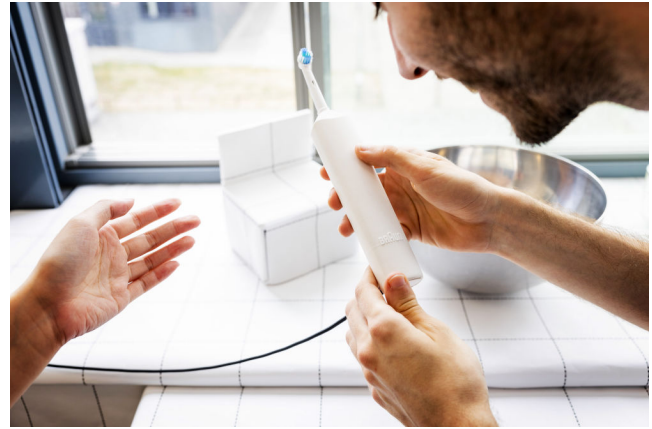

**Figure 15:** *Future Dialogue*: playful control of small home appliances through training on personal vocal languages.

The category *play with AI* includes a single project, *Future Dialogue*, yet we believe it brings an important approach to the design of AI things, that is distinctively designerly. It also has a unique status in the taxonomy, as it involves both disenchanting and re-enchanting. The characteristic element of this approach is the focus on the playful - and to some extent purposeless - engagement with AI. With play, we describe interactions that engage the user in intrinsically motivating and curiosity-driven experiences of AI. Playful approaches to technology are seen as a valuable way to let people explore possibilities, see how new technologies function, and imagine what they can be used for [48]. When allowed to purposelessly interact with AI and 'see what happens', users can experience many of the characterizing elements of play, such as anticipation, surprise, pleasure, and understanding [34]. By playing with technology, the experience of AI products favors the occurrence of 'a-ha' moments that help the user understand the technology, as well as to keep the engagement high.

The *Future Dialogue* project (Figure 15) focuses on a novel mode of interaction with small home appliances. Devices can be controlled only through personalized sounds, and each object first needs to be trained by the user. The underlying speculative narrative is that AI will enable an unprecedented level of personalization in product experiences, including hyper-personalized product interfaces that leverage secret languages defined by users. To engage with this vision, the appliances in the project (a coffeemaker, a toothbrush, and a blender) are provided with components that materialize the key conceptual elements of a machine learning algorithm for classification: a rotary encoder and LEDs are used as an interface for selecting a training category, a button is used to activate the recording of sounds for training, and a switch allows to shift from a training modality to play with the trained objects. This moves the activity of training AI models from an engineering practice to an end-user engagement activity, developed by making use of the MLTK01 board [78]. Through this interface, the objects tell the story of a possible future but also open the space for playing and testing the limits of the AI. This disenchants the users who grasp the hidden workings of AI, yet, at the same time also invites the user to suspend their disbelief and "play along" [99] with the AI products. This allows for reflective cycles of enchantment and disenchantment as play uncovers new qualities of the underlying mechanisms which are quickly deployed by the end user as new interactions, where the mechanism disappears again.

This design approach, then, playfully engages with distinctive AI characteristics, such as unpredictability and uncertainties [19], that when encountered in interfaces could disenchant the user [99], and intentionally makes them part of the experience. This creates new spaces for enchantment where we don't trick ourselves and the audience into believing in the magical and supernatural powers of AI, but rather invite them and us to suspend disbelief and play along with the pretense that AI things work as if they were magical.

## 6.7 Presume AI

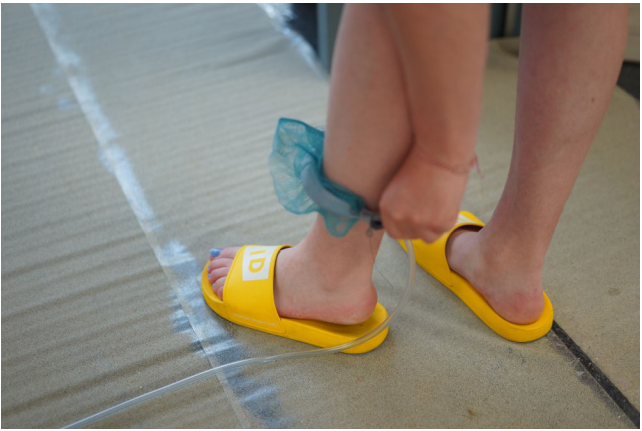

**Figure 16:** *Reflection through Collection*: **data rituals for surfers engaged in citizen science**

This category includes projects (4) that do not engage explicitly with AI, but design products that live within the data-driven world of AI and implicitly provide anchors for imagining potential connections with AI and ML capabilities. The *Reflection through Collection* project (Figure 16), for instance, focuses on the development of a leash for surfers that has sensors embedded for collecting water quality data, such as temperature, PH levels, salt quantities, and more. The whole concept revolves around the ritual of wearing and then washing the leash as a way to first collect and then release data. The act of 'releasing' data implicitly suggests the presence of an intelligent processing element in the system that does something with that data, yet this remains unaddressed and something for the audience to wonder about. In this category of projects, we see design presuming AI as a background condition for the project to exist, although this does not come directly into the interactions with the products and services envisioned. In contrast to other categories, here design does not explicitly operate on a specific direction (dis)enchantment. A product can either be enchanting or disenchanting and simultaneously use other principles, such as magic metaphors or the manifestation of mechanisms. In *Reflection through Collection*, for instance, there is no explicit AI enchantment at play – although one could argue for the project drawing on *Summoning of AI as supernatural entity* – but the ritual of washing out data from the leash clearly sets out to create an enchanting experience of the wearable technology.

## 7 DISCUSSION

The descriptions of each category in the *(un)making AI magic* taxonomy provided here illustrate different strategies by which designers can act within the space of AI and magic. In this discussion, we look at the relations and dynamics between the strategies and the design projects, discuss the limitations of this work, and then discuss the implications of the taxonomy for both design/HCI research and practice.

### 7.1 Taxonomy dynamics

The taxonomy unpacked here is not designed to precisely and uniquely define the position of particular projects, but rather to help navigate the space of AI magic by articulating various principles and their effects, to support the responsibility and agency of designers who deploy them. Here, we draw out some key relations between projects, designers, and the principles we have derived.

First of all, *each project can embody multiple principles*. For instance, the artifact in *Under the Loop* project (Figure 17) *applies magic metaphors* of enchanted magnifying glasses (e.g. [100]), declaring the intent of scrutinizing space for invisible signals. At the same time, the description of the functioning also reveals that the designers presumed a layer of connectivity and algorithmic processing (*presume AI*) and that there is a potentially disenchanting role to play as the *mechanisms are manifest* and 'citizens might be prompted to provide input on specific locations'. All in all, however, the smoothness of the designed artifact and related interaction, together with the lack of technical details in the description of the project manifests an emphasis on the enchanting power of the project, an interest in creating a seamless experience, for which technical mechanisms remain hidden.

The *Colored Realities* project, which we used to exemplify the principle of materializing beliefs, may also be seen as a reinterpretation of a magical thing or a sophisticated transposition of stage magic principles. The dynamically changing configuration of the news piece in the project is reminiscent of the *'live newspaper'* from the world of Harry Potter in which pictures become alive when the reader is watching, but also represents a seamless embedding of the stage magic principle of transformation that feels enchanting to the user who experiences it. Similarly, the *A Closer Look* project, which we used to describe the principle of manifesting mechanisms, clearly also borrows the metaphor of the magic mirror and affords *play with the AI* algorithms, by providing buttons and sliders that instantly let the user affect the product behavior. However, the overall emphasis is put on the intricacies of the mechanisms at play and reveals an explicitly critical view about AI.

*The effects of combined principles in a single project tend to either enhance or diminish the perception of the magical aura of AI.*

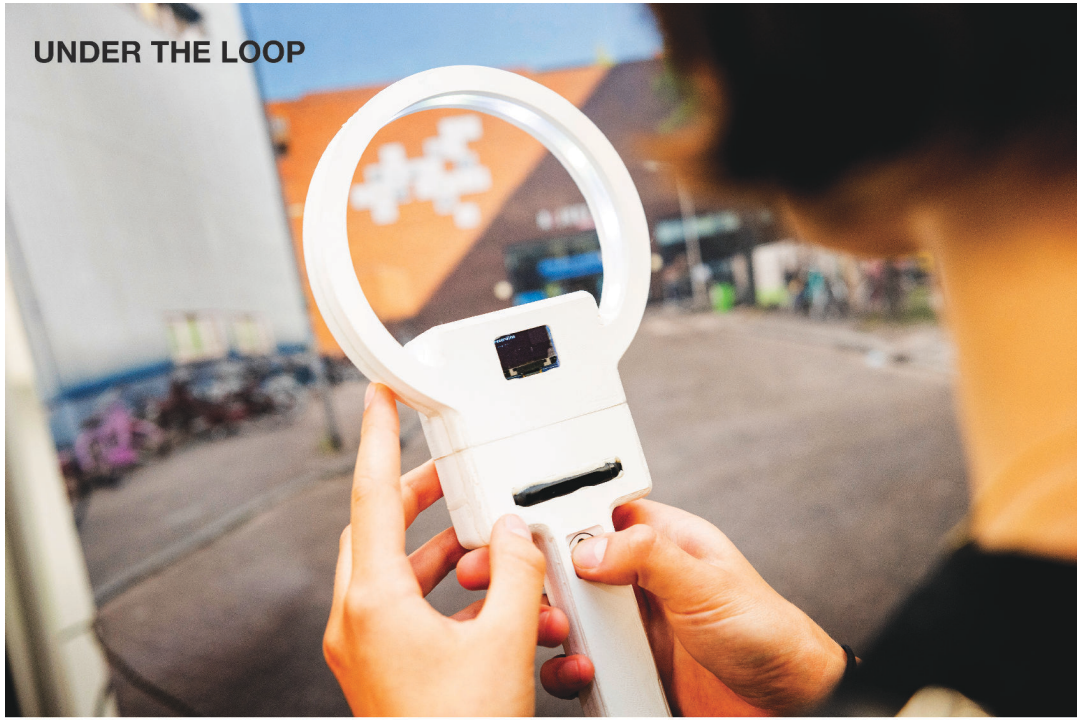

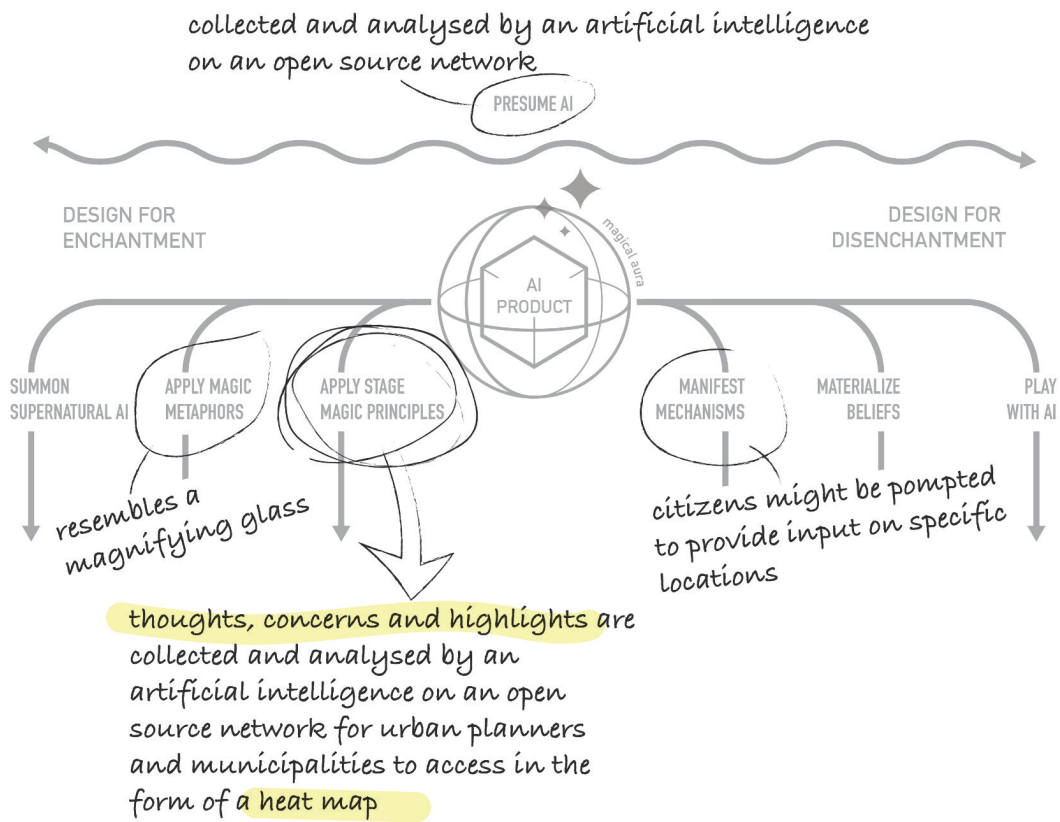

Figure 17: Analysis of *Under the Loop* project. The example shows how a single project and its description can map onto different principles. However, one principle and an effect (enchantment or disenchantment) tend to prevail.

While multiple principles can coexist in a single project, in most of the projects one dominates over the others. This means that if a designer wishes to increase or decrease enchantment, the taxonomy offers multiple complimentary routes to do so: the principles give inspiration for how to design enchanting or disenchanting AI products, and consciously direct the effects that those products will have on user's perception. The exception to this is *play with AI*, which asks for a continual iteration of enchantment and disenchantment.

The taxonomy can be used as a reflexive tool to understand one's attitude and standing and to shape the choice of whether critical or affirmative principles are manifested most prominently. However, when enchanting and disenchanting principles are both present the effects of disenchanting principles tend to dominate. In the projects here, *materializing beliefs* and *manifesting mechanisms* take over, as we "open up the unfamiliar device to see what makes it tick [and] our reductionist curiosity and microanalysis kill the enchantment" [104].

*These principles affect the designers as well as the intended end users*. Through our analysis, it was clear that the magical aura of AI affects not only the user but also, and foremost the designer. The principles of *summoning AI as magical entity*, *manifesting mechanisms* and *materializing beliefs* in particular, put the spotlight on the designers themselves, and their fascination or fears about AI. The designers could hold multiple standpoints through their projects; for example, the *manifesting mechanisms* and *materializing beliefs* principles declare a critical view of AI, yet this does not prevent the designers from deeply engaging with the technology. In contrast, the projects embodying the *summoning AI as magical entity* principle often did not deeply engage with the AI tools and technical components, instead, AI is rather 'talked about' and staged through Wizard of Oz techniques. This may sound counter-intuitive as one would think that fascination for AI may lead to a rich exploration of it, while skepticism may discourage engagement. However, this tends to have to do with the actual level of engagement and making with AI, and the corresponding disenchantment on the part of the designers. The *play with AI* principle is again an exception as it works both to disenchant and re-enchant and invites the user to play along with the designer. Users can participate in a continuous reconfiguration of AI that brings awareness on the functionings and limits of the technology while feeling the magical aura when this performs as desired.

Finally, *a rich engagement with AI is key to developing a conscious approach to the design of AI products*. Designers in this survey who had a strong engagement with AI were able to leverage the dynamics of magical thinking and doing somewhat purposefully, to develop their positionality. This does not mean that designers should necessarily express a critical stand towards AI and only design for disenchantment but rather be aware of the position they take and be responsible for the effects they might generate on users through their designs. In contrast, the designers who engaged less strongly with the workings of the AI technology they were using, found it harder to construct an articulated position, whether critical or affirmative.

## 7.2 The Taxonomy as a Nascent Design Theory of AI Magic

To further appreciate the value of the *(un)Making AI Magic* taxonomy, we should consider how this responds to the HCI need for more ontological resources [23, 75], and specifically around the space of AI perception and design [31].

The taxonomy, we argue, can be considered as a form of theory in the broad sense [55], what Forlizzi terms *nascent theory* [45]: it suggests a lens through which we can understand phenomena and dynamics around the design of AI things and stimulate questions by introducing new constructs and proposing relations between these and established concepts. The design principles composing our taxonomy (illustrated in Section 6) and the related dynamics (Section 7.1) introduce new constructs.

This lens allows us to look at the existing research landscape around AI, in particular connecting creative research endeavors with more technical approaches. For instance, *Manifesting Mechanisms*, a disenchanting principle, has strong analogues in technical approaches such as algorithmic auditing [67] and explainable AI approaches [5], but also connects to the critical and speculative explorations we saw in students' projects. *Apply Magic Metaphors*, instead, intuitively connects to artistic and designerly research endeavors in the space of metaphors and product semantics, such Benjamin et al.'s Entoptic Field Camera [11] that looks at how a perceptual concept can help us understand AI, or Murray-Rust et al. [91] who explored how metaphors can help designers in thinking about the creation of AI systems. However, the framing in magic metaphors also encourages investigations into matters of human cognition, and mental models in particular.

The *(un)Making AI Magic* taxonomy, then, provides anchors for different HCI research identities to engage with, and opportunities for cross-pollination. Nevertheless, we argue for this to be primarily a Design theory (as opposed to a theory for design) [103] since it foregrounds making and experiential exploration as a primary way for understanding AI and magic. It specifically *centers around the exploration of multiple ways there can be to make AI* and the implications of these possibilities in terms of enchantment's intentions and effects.

Due to this centering on making as a way of learning about AI, in what follows, we unpack potential implications of the taxonomy for both design education and wider practice.

## 7.3 How to (un)Make AI Magic

Taxonomies are powerful theoretical instruments that can strongly impact future research, as demonstrated by popular examples, such as the *Taxonomy for Autonomous Agents* [46] which is now a foundation for thinking about artificial agents and multi-agent systems, or the *Socially Interactive Robots* taxonomy [44], that has become a key resource for human-robot interaction designers. Taxonomies are useful for several purposes, from the classification of existing knowledge to the identification of knowledge gaps, from the identification of objects and characteristics to the positioning of a research output, and more [110]. Despite successful examples, however, the HCI field historically suffers from a gap between research and professional design practices, and even more, between theory and specific design instances [120].

Table 1: Translating principles into 'what if' questions to use the taxonomy for design exploration

| Principle | What if... |
| --- | --- |
| Presume AI | ...product X would implicitly leverage an AI infrastructure? |
| Summon Supernatural AI | ...product X would have supernatural AI capabilities? |
| Apply Magic Metaphors | ...product X would embed AI looking and behaving as Y? |
| Apply Stage Magic Principle | ...product X would use AI as a magic trick? |
| Manifest Mechanism | ...product X would declare the AI mechanisms embedded into it? |
| Materialize Beliefs | ...product X would manifest designers' or users' beliefs about AI? |
| Play with AI | ...product X would invite users to play with AI? |

In this vein, our ambition here is to create a taxonomy that can be applied and appropriated in many situations. We suggest two particular ways to make use of the *(un)Making AI Magic* taxonomy for design/HCI practice: *exploration* and *reflexivity*.

*7.3.1 'What if?' – using enchantment for design exploration.* The seven principles composing the taxonomy can be explicitly used for inspiring and guiding design interventions by translating each into a 'what if' question (see Table 1) to support exploratory design processes [109].

This simple act of translation from principles to *what if* questions is highly generative as it automatically opens the way to a variety of practical questions. In Figure 18 we illustrate this using the project *Glow* (project 24, year 2), through an exploration of the questions from Table 1.

**A What if the product leverages an AI infrastructure?** In this initial framing, the artifact does not embed any AI capability but clearly lives in a data-driven world *presuming hypothetical AI* capabilities to function, warming up when the user needs it based on contextual and user data.

**B What if it has supernatural AI capabilities?** This begs the question *what kind of superior ability related to heating and sensing the environment would be sufficiently sophisticated to make it perceived as 'supernatural'*. Based on existing possibilities, one could imagine a product able to 'see' through walls and eventually turn on.

**D What if the product uses AI as a magic trick?** invites us to consider *what spaces for unexpectedness there can be*. What would be a surprising and counterintuitive modality of interaction? The answers can be many. One could be functional anticipation of the user's state and need, e.g., the heater predicting that the user would feel cold after working on a laptop for a while and heating up preventively. Another could be interactional, where AI is leveraged to detect users' gestures for activation.

**C What if the heater was a magical creature?** Multiple metaphors can be used, but only some can 'talk' about both the functional scope of the product as well as the capabilities of AI embedded into it. The magical figure of a *phoenix*, for instance, could be used as a metaphor for designing a heater that uses AI capabilities to 'see' user need for heat (as in *Summon Supernatural AI*), but that would completely erase – or burn – any data and knowledge about the user any time it 'dies' – or f-switches off – and is reborn as a way to preserve privacy.

**E what if the heater would declare the AI mechanisms embedded into it?** asks foremost to reflect on what the heater's mechanisms and their implications are, drawing on explainable AI techniques to either aid user understanding or highlight privacy concerns as a result of hyper-personalization. For the latter, one could design a heather that shows how a user is being tracked and provoke a reflection on how we grant smart products access to personal data.

**F What if the heater manifests beliefs about AI heaters?** One could assume a user's belief of AI as objective and infallible, and hence better suited to making decisions regarding sustainable heating behaviors. The heater could then be designed to switch on exclusively when 'objectively needed' with no possibilities for the user to control, resulting in a frustrating and confrontational AI heating interface.

**G What if the heater would invite users to play with AI?** This could potentially touch on many of the design concepts mentioned above, such as gesture detection, environment data collection, the manifestation of user monitoring, and more, but with the explicit intent of enabling exploration playfully. Here the emerging questions would be about how a heater can be an interface for interacting with AI and *what affordances can we design to enable play*.

As this brief thought exercise shows, the taxonomy principles when turned into *what if* questions open a generative design space. These allow us to consider a wide spectrum of possibilities and offer a framing for the confrontation of design ideas that can also help detect possible issues in the conceptualization of AI products, such as design fixation [63] and normativity [47].

*7.3.2 'Why?' - investigating enchantment for reflexivity in design.* The taxonomy can also be used to encourage designers to reflect on their intentions and positionality. *What if* questions come with effects in terms of enchantment, but *Why?* questions have implications in terms of crticiality in design practices [6]. As Fallman [40] argues, through exploration, design can show alternatives and examples, but it also becomes a statement of what is possible, what would be desirable or ideal. The principles in the taxonomy can pull in different directions: *Apply Stage Magic Principles* enchants through seamless interaction while *Manifest Mechanisms* disenchants through confrontation the second.

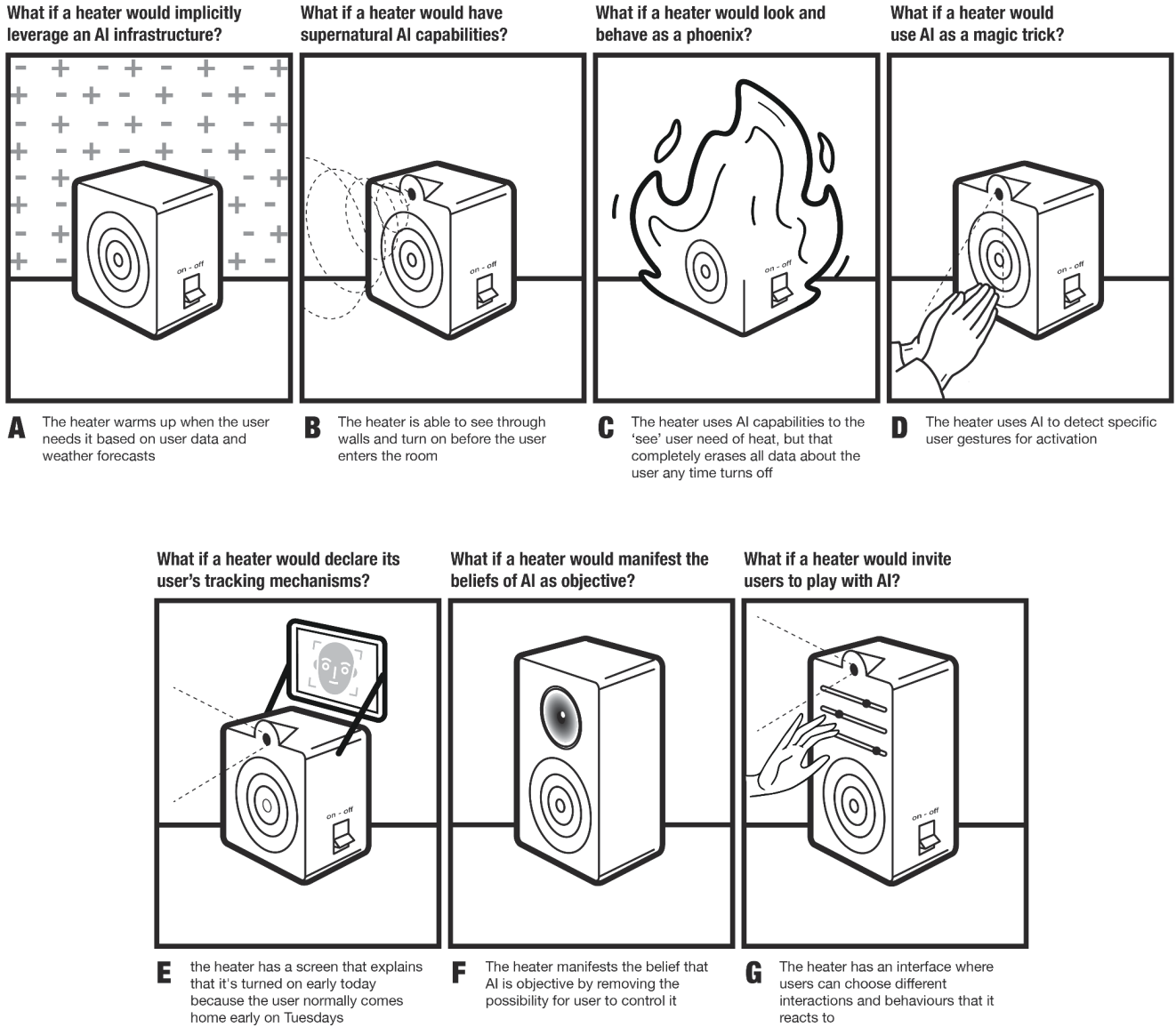

Figure 18: Examples of 'what if questions in use: What if a heater would...

These potentially conflicting viewpoints can serve as a frame for reflexivity and scrutiny if looked at as a 'map' for designers to reflect on their positionality. As background, there are many examples from professional design work that materialize the principles included in the taxonomy. AI products integrate metaphors from the world of magic, such as the *The Mirror Home Gym* that uses the metaphor of the magic mirror, or the *Bosch Series 8 Accent Line Sensor Oven* [13] that uses the metaphor of the crystal ball. Others use stage magic principles to create smooth and enchanting user experiences, such as the family of Amazon products where *Alexa* manifests the principle of teleportation by giving the impression of 'jumping' from object to object, or the *NEST* smart thermostat that learns and estimates house heating behaviors [62] to predict user needs and anticipate actions. In these examples, there is hardly any engagement with matters of values, assumptions, and power which, instead, are core to reflexive practices [97].

In contrast, examples from critical and speculative practices [127] inherently come with a more explicit commitment to the critique of norms [6]. For instance, the design studio *Superflux* confronts us with the beliefs we imbue technology with, such as *Our Friends Electric* [115] which explores the assumptions we hold regarding our relationship with voice-activated AI assistants.

The *Paragraphica* camera [121] uses location data and AI to visualize a "photo" of a specific place and moment – an explicit reflection on aspects of data processing and decision-making around what is deemed relevant and, thus, 'captured' from the environment: a clear example of the principle of *Manifesting Mechanisms*. Reflexivity in these cases usually revolves around the development of a single standing, the materialization of a personal view, rather than the exploration of a spectrum of possibilities.

Table 2: Translating principles into 'why' questions to use the taxonomy for design reflexivity

| Principle | Why... |
| --- | --- |
| Presume AI | ...would a product X implicitly leverage an AI infrastructure? |
| Summon Supernatural AI | ...would a product X have supernatural AI capabilities? |
| Apply Magic Metaphors | ...would a product X embed AI looking and behaving as Y? |
| Apply Stage Magic Principle | ...would a product X use AI as a magic trick? |
| Manifest Mechanism | ...would a product X declare the AI mechanisms embedded into it? |
| Materialize Beliefs | ...would a product X manifest designers' or users' beliefs about AI? |
| Play with AI | ...would a product X invite users to play with AI? |

To facilitate the use of the taxonomy for reflexive purposes, we suggest another simple act of translation: from principles to 'Why?' questions (Table 2). In asking why, we automatically engage with matters of intention, effects, and values that an AI artifact can carry.

Looking back at the example of the *smart desk heater*, we could then ask *why a desk heater would be able to see through walls*. Answering this would invite us to engage with matters of control and agency, as such capabilities enable the product to make decisions and act autonomously. Or one could ask *why a desk heater would behave as a phoenix*, leading us to consider matters of data protection and privacy. Such 'Why?' questions should be engaged not only for reflecting on what is being designed but also and foremost for considering how else AI things could be shaped. Asking *why* then helps us consider *what are the underlying values and narratives* we are embedding into AI products, and *what alternative stories we could wave*. The taxonomy provides a platform for diverging from either norms or personal views, opening up a spectrum of alternatives.

### 7.4 Limitations and Future Work

Our work, of course, is not free from limitations. In particular, one could question how relevant a taxonomy is for professional design practices when it's built out of reflections on students' design work. Several are the differences in the way students approach design as compared to professionals. For instance, experienced designers – in contrast to students – have tendencies to refer to past designs, to question if a project is worth pursuing, and to consider possible issues [3]. Furthermore, students' work is highly influenced by the perspectives and practices brought into the course by the educators. In our collection, for instance, we see a connection between our own critical and experimental approach to design research and education, and the students' projects. Design theories, however, are often built by reflecting on educational work (see more in Section 5). Students' work is valuable not because of its potential capacity to emulate professional design practices, but rather because of its dialogical relationship between the educators' conceptual understanding of a theoretical space surrounding design/HCI practice and its manifestations. It is exactly in this dialogical space between the educators' standing and the students' interpretation that we find a powerful ground for confrontation and reflection, where we 'retroactively' look at our education practice to understand how theoretical principles can translate and be appropriated into practice [103].

We acknowledge, however, that as for most design theories, our taxonomy is generative and suggestive, rather than verified [50]. Further empirical evidence could be gained both in the classroom and in professional environments by using the taxonomy as a guide for AI product development or as a reflexive tool for self-scrutinizing processes and then running comparative studies. Future empirical investigations could look for unpacking the dynamics of enchantment and disenchantment: How does a designer's experience affect the way they use the principles? How does the context of a project affect the success of applying any of these techniques? How does prolonged exposure for designers or end users affect the perception of the magical aura? For all of these dynamics, we found hints in our work, yet they open up a broad space that could not be addressed in this single analysis. Nevertheless, we as design/HCI researchers should be mindful of the role of the different ways of knowing that exist in the HCI landscape and that design theory, compared to theory for design, is hardly generalizable and verifiable, but nonetheless holds great value in its capacity to inspire future work and drive research agendas [50, 103].

Thus, the taxonomy presented here is open-ended and represents a first–high-level–layer classification in the space of *(un)making AI magic* which could be seen as a limitation, but we look at this optimistically: as Nickerson and colleagues [95] argue "a useful taxonomy should allow for the inclusion of additional dimensions and new characteristics". For each design principle, additional characterizations can be found and nuances in the related dynamics can be understood. For instance, the general principle of using magic metaphors can be further investigated to understand how different magic categories, whether creatures, objects, or 'fait' types [17], can bring distinct implications for how a product is perceived and approached. Similarly, one could develop specific practical techniques for working with each of the principles, for instance by delving into the distinct effects and implications of diverse stage magic techniques. We hope that this initial framing gives a fertile structure for building such future work and that more design/HCI researchers will feel encouraged to engage in theory-making work.

## 8 CONCLUSIONS

In this paper, we illustrated and problematized the fuzzy design space of magic and enchantment in AI product development. Designing enchanting products contributes to developing seamless and engaging experiences, but can also encourage users' misunderstanding of products' capabilities and hide the complexity and risks of ecosystems populated by AI products. Designers need to navigate between the ambition to be critical and conscious about the use of AI while also being caught up in the magic rhetorics that drive the 'economy of appearances' [119] shaping public engagements. Enchantment is not just designed but is inherently ingrained in novel technologies as it is tightly bound to the difficulty of grasping the mechanisms at play: roads and railways once enchanted with their stories of speed and freedom [58], now AI technologies enchant with their dreams of agency and care. Designers hold a responsibility to approach AI technologies carefully and skillfully. For this, it is of primary importance to build literacy around AI not only in its technical terms but also in its socio-cultural components. Hence, it is crucial to understand the more or less subconscious principles of magic and supernatural thinking that heavily characterize current AI development, in order to shape the societal and cultural impact of these technologies.

As the field still *lacks an understanding of the dynamics and effects of magical thinking in the design of AI things*, we developed a taxonomy of the different configurations of magic present in the design of AI things. By reflecting primarily on student design projects but also connecting professional design works, we identified and unpacked the implications of 7 design principles that can distinctively characterize AI products' development: *applying stage magic principles*, *applying magic metaphors*, *summoning AI as supernatural entity*, *materializing beliefs*, *manifesting mechanisms*, *play with AI*, and *presuming AI*. These, we believe, provide an initial overview of different ways design can operate in relation to AI and emphasize or diminish its enchanting power.

We do not see this taxonomy as closed, and hope that the 'unfinished' nature of our taxonomy serves as an invite for the community to engage and contribute. It is a framework to look at AI magic not as a marginal and disconnected topic but as a lens to reflect upon and understand many of the critical perspectives that are currently being investigated in the design and HCI field to make sense of AI, such as metaphors [8, 11, 70, 91], narratives [20, 56, 60, 77], and expressive implications of AI uncertainties and errors [10, 19].

To conclude, we invite designers and researchers finding themselves engaged in AI products' development to reflect upon their thinking and whether there is some 'magic at play'. For this, we offer at our taxonomy as a reflexive tool that allows critical scrutiny of the things we design and the expectations we bring in our actions. As such, we believe our work will contribute to building a disciplinary sensitivity for the role of rhetorics and 'the irrational' in the design of AI things, and technology more broadly.

## ACKNOWLEDGMENTS

This work owes a great deal of gratitude to all the students who passionately engaged with the Interactive Technology Design course at TU Delft. Their ways of appropriating the technologies and the AI discourse were an inspiration and a motivation for us as educators to grow awareness about aspects of AI perception and rhetoric. Especially, we would like to thank Lenny Martinez who relentlessly contributed to this work as a design student first, and as a teaching assistant for the course later.